\documentclass[11pt,preprint]{article}
\pdfoutput=1

\usepackage{hyperref}
\hypersetup{
  colorlinks=true,
  linkcolor=darkpurple,
  citecolor=greenLinks,
  urlcolor=redLinks,
  pdfborder={0 0 1}
}

\usepackage{lmodern}
\usepackage{booktabs}
\usepackage[T1]{fontenc}
\usepackage[latin9]{inputenc}
\usepackage{slashed}     %\slashed{p}
\usepackage{bbold}  % \mathbb{1} for the identity matrix
\usepackage{url}
\usepackage{graphicx}
\usepackage{float}
\usepackage{color}
\usepackage{multirow}
\usepackage{amsmath}
\usepackage{amssymb}
\usepackage{breqn}
\usepackage{ulem}
\usepackage{authblk}
\usepackage{mciteplus}
\usepackage[numbers,sort&compress]{natbib}
\usepackage[left=2.55cm, right=2.55cm, top=2.55cm, bottom=2.55cm]{geometry}

\definecolor{darkred}{rgb}{0.6,0,0}
\definecolor{darkpurple}{rgb}{0.5,0,0.5}
\definecolor{greenLinks}{rgb}{0, 0.6, 0} 
\definecolor{redLinks}{rgb}{0.6, 0, 0}
\definecolor{journalLinks}{rgb}{0.6, 0, 0}
\definecolor{eprintLinks}{rgb}{0.4, 0.4, 0.4}

\def\gsim{\raise0.3ex\hbox{$\;>$\kern-0.75em\raise-1.1ex\hbox{$\sim\;$}}}\def\lsim{\raise0.3ex\hbox{$\;<$\kern-0.75em\raise-1.1ex\hbox{$\sim\;$}}}

\newcommand{\fig}[1]{figure~\ref{#1}}

\newcommand{\eq}[1]{\eqref{#1}}

\begin{document}

%%%%%%%%%%%%%%%%%%%%%%%%%%%%%%%%%%%%%%%%%%%%%%%%%%%%%%%%%%%%%%%%%%%%%%%%%%%%%%%%%%%%%%%%%%%%%%%%%%%%
% TITLE
\title{{\Large{}\vspace{-1.0cm}} \hfill {\normalsize{}IFIC/21-35}\\*[10mm]
  {\huge{}Temperature effects on the $Z_2$ symmetry breaking\\ in the scotogenic model}{\Large{}\vspace{0.5cm}}}
\date{}

%%%%%%%%%%%%%%%%%%%%%%%%%%%%%%%%%%%%%%%%%%%%%%%%%%%%%%%%%%%%%%%%%%%%%%%%%%%%%%%%%%%%%%%%%%%%%%%%%%%%
% AUTHORS
\author[1]{{\large{}Alexandre Alvarez}\thanks{E-mail: \href{mailto:alexandre.alvarez@physik.uni-wuerzburg.de}{alexandre.alvarez@physik.uni-wuerzburg.de}}}
\author[1,2]{{\large{}Ricardo Cepedello}\thanks{E-mail: \href{mailto:ricardo.cepedello@physik.uni-wuerzburg.de}{ricardo.cepedello@physik.uni-wuerzburg.de}}}
\author[2]{{\large{}Martin Hirsch}\thanks{E-mail: \href{mailto:mahirsch@ific.uv.es}{mahirsch@ific.uv.es}}}
\author[1]{{\large{}Werner Porod}\thanks{E-mail: \href{mailto:porod@physik.uni-wuerzburg.de}{porod@physik.uni-wuerzburg.de}}}

%%%%%%%%%%%%%%%%%%%%%%%%%%%%%%%%%%%%%%%%%%%%%%%%%%%%%%%%%%%%%%%%%%%%%%%%%%%%%%%%%%%%%%%%%%%%%%%%%%%%
% AFFILATIONS
\affil[1]{\small Institut f\"ur Theoretische Physik und Astrophysik, \protect\\
  University of W\"urzburg,  Campus Hubland Nord, D-97074 W\"urzburg, Germany \vspace{0.2cm}}

\affil[2]{\small AHEP Group, Instituto de F\'isica Corpuscular, CSIC - Universitat de Val\`encia \protect\\
	Edificio de Institutos de Paterna, Apartado 22085, E--46071 Val\`encia,
	Spain}

\maketitle

%%%%%%%%%%%%%%%%%%%%%%%%%%%%%%%%%%%%%%%%%%%%%%%%%%%%%%%%%%%%%%%%%%%%%%%%%%%%%%%%%%%%%%%%%%%%%%%%%%%%
% ABSTRACT
\begin{abstract}

It is well-known that the scotogenic model for neutrino mass
generation can explain correctly the relic abundance of cold dark
matter. There have been claims in the literature that an important
part of the parameter space of the simplest scotogentic model can be
constrained by the requirement that no $Z_2$-breaking must occur in
the early universe. Here we show that this requirement does not give
any constraints on the underlying parameter space at least in those 
parts, where we can trust perturbation theory. To demonstrate this, we
have taken into account the proper decoupling of heavy degrees of
freedom in both, the thermal potential and in the RGE evolution. 
  
\end{abstract}

%%%%%%%%%%%%%%%%%%%%%%%%%%%%%%%%%%%%%%%%%%%%%%%%%%%%%%%%%%%%%%%%%%%%%%%%%%%%%%%%%%%%%%%%%%%%%%%%%%%%
% MAIN BODY
\section{Introduction}

Among the known shortcomings of the standard model (SM), probably dark
matter (DM) and neutrino masses are the most important ones. While
many extensions of the SM have been proposed that could either explain
DM or neutrino masses, the ``scotogenic'' model \cite{Ma:2006km} can
be considered the proto-type for a class of beyond SM (BSM) models
that can potentially explain both phenomena at the same time. 

\begin{figure}[t]
\centering
\includegraphics[width=0.5\textwidth]{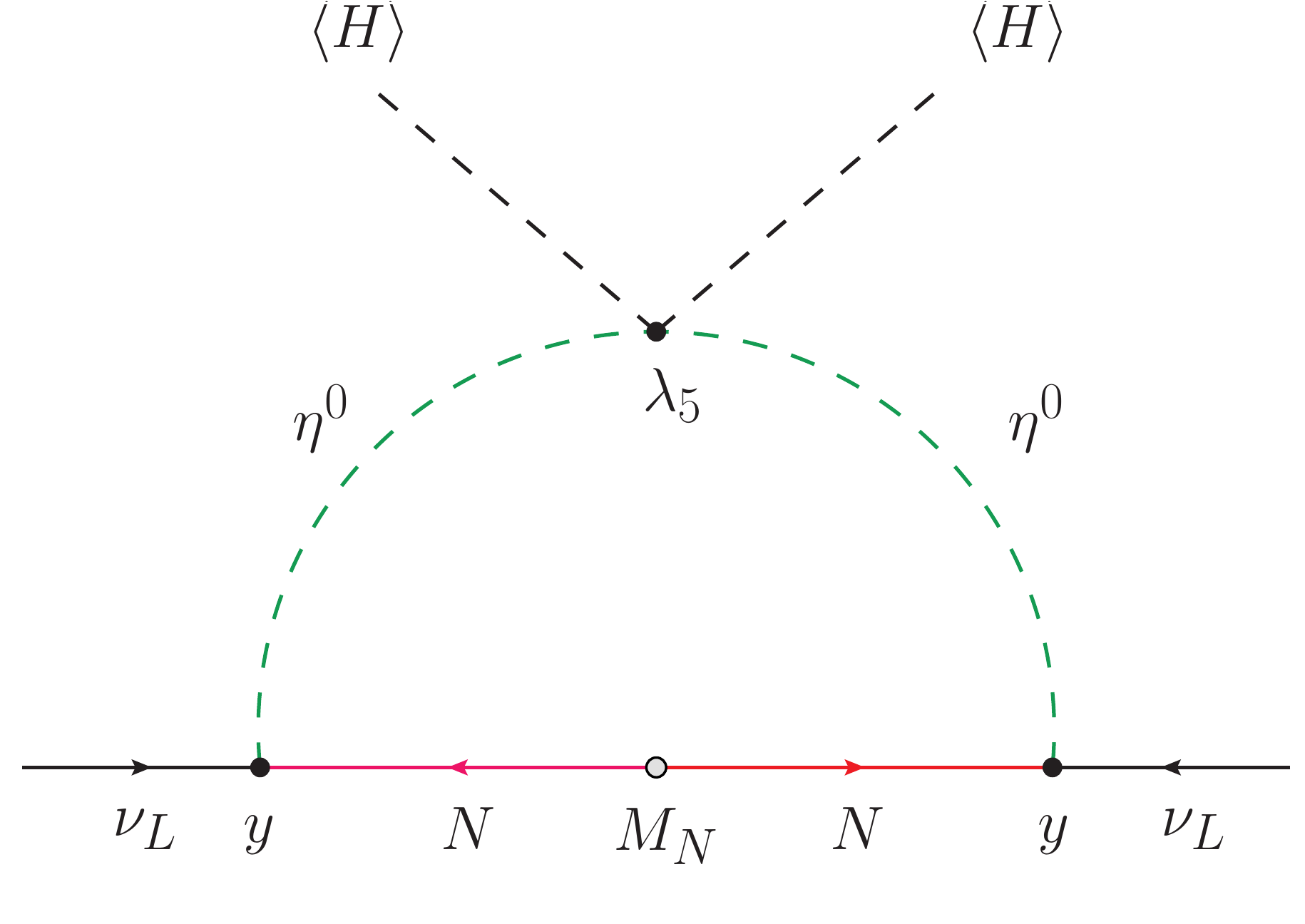}
\caption{One-loop neutrino mass diagram in the scotogenic model.}
\label{fig:SC}
\end{figure}

The setup of the scotogenic model is very simple: Add
to the SM (two or) three right-handed neutrinos, $N_i$, and a new scalar doublet,
usually denoted by $\eta$. In order to avoid a tree-level type-I
seesaw, a global $Z_2$ is added by hand. This symmetry eliminates a
certain term from the scalar potential and avoids a vacuum
expectation value for $\eta$, as long as certain conditions on the
scalar parameters are fulfilled, see  section \ref{sect:mod}.  Under this symmetry, $N_i$ and
$\eta$ are odd, while all SM particles are even. The lightest $Z_2$
odd particle, either $N_1$ or $\eta$, is then absolutely stable,
providing potentially a good cold dark matter candidate.  At the same
time, the model is able to explain correctly the observed neutrino
masses and mixing angles (for a recent global fit of all neutrino data
see, for example \cite{deSalas:2020pgw}) via the one-loop diagram shown
in \fig{fig:SC}.

Due to its economic structure the scotogenic model has been studied in
many publications in the literature. Both DM candidates, the lightest
$N_1$ \cite{Suematsu:2009ww} and the the neutral component of $\eta$
\cite{LopezHonorez:2006gr}, can give a relic density consistent with DM
abundance in viable parts of parameter space, while respecting bounds
from direct \cite{Guo:2018iix,Schmidt:2012yg} and indirect
\cite{deBoer:2021pon,deBoer:2021xjs} detection.

On the other hand, for any model to generate the correct dark matter
abundance observed today requires the DM candidate to be stable from
the time of its decoupling from the thermal bath in the early universe
onwards. The scotogenic model is no exception: Breaking the $Z_2$ at
this -- or any later -- time would lead to a rapid decay of the DM
candidate and consequently, both the DM and the motivation to study
the model would be lost.  The authors of \cite{Merle:2015gea} have 
pointed out that when evolving the parameters of
the scotogenic model to larger energies using RGEs, very often
points which had a conserved $Z_2$ at low energy develop a
deeper minimum at a non-zero value of the vev of $\eta$ at high
energies and, thus, $Z_2$ is spontaneously broken. This is induced by
a negative contribution arising from a one-loop correction to the mass
of the $Z_2$-odd scalar doublet mediated by the $Z_2$-odd Majorana
right-handed heavy neutrinos, $N$.  It was then argued in
\cite{Merle:2015gea}, that all such points should be excluded from the
parameter space for consistency.

However, the early universe is a hot thermal bath and temperature
effects, while briefly discussed, were not taken into account in a
consistent manner in \cite{Merle:2015gea}. It is well-known that the 
addition of thermal effects is responsible of the electroweak symmetry 
restoration at temperatures around $165$ GeV \cite{Dine:1992wr,Katz:2014bha}.
One then would expect that thermal contributions should mitigate or 
even avoid these $Z_2$ breaking minima at high scales. In this paper, we discuss
how to build the effective potential of the scotogenic model
\cite{Ma:2006km} including one-loop thermal and non-thermal effects,
as well as the corresponding resummation terms. We then
\textit{improved} the potential by taking into account the RGE
evolution of the parameters, in order to consider the possibility that
the $Z_2$ is spontaneously broken at a high energy scale, which is characterized by the temperature $T$. As a result,
we find that the inclusion of thermal effects change the conclusion about
the stability of the DM in the early universe.  With the possible
exception of parameter points close to non-perturbativity, where we
can not trust our calculation, we found no points in our scans in which
the $Z_2$ is broken spontaneously. In particular, our conclusions
hold in the parameter region studied in \cite{Merle:2015gea}.
\\

The rest of this paper is organized as follows. In section \ref{sect:mod}
we give a description of the scotogenic model and its parameters. We
also discuss briefly the tree-level stability conditions and the constraints these
put on the parameter choices for the scalar potential. Section
\ref{sect:eff} is devoted to a discussion of the effective potential,
temperature effects and the correct decoupling of the heavy degrees of
freedom. In section \ref{sect:num} we discuss our numerical results.
The paper then closes with a short summary.

\section{The scotogenic model\label{sect:mod}}

This section gives a brief description of the scotogenic model and
a discussion of its scalar potential. The model is a simple extension
of the SM, to which three copies of right-handed neutrinos, $N \propto
(1,1,0)$, and a new scalar doublet, $\eta \propto (1,2,1/2)$, are added.
Here, $(x,y,z)$ indicate quantum numbers under the SM group in the usual
order $SU(3)_C\times SU(2)_L \times U(1)_Y$. $N_i$ and $\eta$ are
assumed odd under a new $Z_2$ symmetry. The Lagrangian of the model
then contains the terms:
\begin{align} \label{eq:LagN}
  \mathcal{L} &\supset - (Y_N)_{\alpha\beta} \, L_\alpha \eta^\dagger N_\beta
    -\frac 12 M_{N_\alpha} N_\alpha N_\alpha -  \mathcal{V} \\
    \mathcal{V} &= 
    \mu_1^2 \, H^\dagger H + \frac 12 \lambda_1 \, (H^\dagger H)^2 + \mu_2^2 \, \eta^\dagger \eta + \frac 12 \lambda_2 \, (\eta^\dagger \eta)^2 
    \nonumber \\
    &+ \lambda_3 \, (H^\dagger H) (\eta^\dagger \eta) + \lambda_4 \, (H^\dagger \eta) (\eta^\dagger H) + \frac 12 \left[ \,\lambda_5\, (H^\dagger \eta)^2 + \text{h.c.} \, \right]\,.
 \label{eq:scalar_pot}
 \end{align}
Here, the fermions are described in terms of 2-component spinors and $H$ is the usual SM Higgs doublet. Note that the $Z_2$ symmetry eliminates terms
such as $m_{12}^2 \, H^{\dagger}\eta$. We will assume in the following that $\lambda_5$ is real, as this does not affect any of our findings.

Neutrino masses arise at one-loop order as shown in \fig{fig:SC}. The $3 \times 3$ Majorana neutrino mass matrix from this
diagram is given by
\begin{align}
\left(m_{\rm SC}\right)_{\alpha\beta} &=
\sum_{i=1}^3\frac{Y_{Ni\alpha}Y_{Ni\beta}}{2(4\pi)^2} M_{N_i}
\left[\frac{m_R^2}{m_R^2-M_{N_i}^2}\log\left(\frac{m_R^2}{M_{N_i}^2}\right)
-\frac{m_I^2}{m_I^2-M_{N_i}^2}\log\left(\frac{m_I^2}{M_{N_i}^2}\right)\right] \, 
\nonumber
\\ &
 \equiv \frac{1}{32 \pi^2} \left( Y_N^T \, {\widehat M} \, Y_N \right)_{\alpha \beta} \, ,
\label{eq:numassSC}
\end{align}
where $m^2_{R/I}$ are the masses of the neutral scalar and pseudoscalar components of $\eta$:
\begin{align}
m^2_R = \mu^2_2 + \frac{1}{2}\left(\lambda_3+\lambda_4+\lambda_5\right) v^2_1 \, , \quad
m^2_I = \mu^2_2 + \frac{1}{2}\left(\lambda_3+\lambda_4-\lambda_5\right) v^2_1 \, ,
\label{eq:masses_eta_R_I}
\end{align}
 with $v_1 = \sqrt{2} \langle H \rangle$. 
Equation~\eqref{eq:numassSC} is of a form which allows to use a minimally
modified version \cite{Cordero-Carrion:2018xre} of the Casa-Ibarra
parametrization \cite{Casas:2001sr} to fit all neutrino data. Since
such fits have been discussed many times in the literature, we will
not repeat any details here. We stress the importance of the term
proportional to $\lambda_5$ in \eq{eq:scalar_pot}. In the limit $\lambda_5\to 0$ the model
conserves lepton number and the Majorana neutrino mass diagram vanishes
identically.\footnote{In more technical terms, one sees from \eq{eq:masses_eta_R_I}
that $\lambda_5=0$ implies $m^2_R=m^2_I$, which in turn gives $\widehat M=0$.} Since this corresponds to an enhanced symmetry, a small
value of $\lambda_5$ is technically natural and somewhat small values of
$\lambda_5$ are usually used in the neutrino mass fit, allowing the
Yukawas to be even order ${\cal O}(1)$. 
\\

We are interested in the question whether $Z_2$ breaking points can
appear during the thermal evolution of the universe, which for
temperature $T=0$ would nevertheless appear to be $Z_2$ symmetric. For
this we have to take into account the thermal contributions to the scalar
potential as discussed in the subsequent section. These corrections
can be expressed in terms of tree-level masses for which we give here
the corresponding formulas allowing for a potential spontaneous $Z_2$
breaking. We decompose the scalar doublets as follows
\begin{equation} \label{eq:Heta}
    H^T = \left( G^+, \; \frac{1}{\sqrt{2}} (v_1 + h + i \,G_0) \right), \qquad \eta^T = \left( \eta^+, \; \frac{1}{\sqrt{2}} (v_2 + \eta_R + i \,\eta_I) \right)\,,
\end{equation}
where we explicitly allow for a possible non-zero vacuum expectation
value for the $Z_2$-odd field $\eta$. We have slightly abused notation here,
as in case of a $Z_2$ breaking the would-be Goldstone fields are an admixture
of $G^+$ and $\eta^+$, as well as an admixture of $G^0$ and $\eta_I$.

We get the background potential by expanding \eq{eq:scalar_pot}:
\begin{equation} %\label{eq:}
    V_0 (v_1, v_2) = \frac 12 \mu_1^2 \, v_1^2 + \frac 12 \mu_2^2 \, v_2^2 + \frac 18 \lambda_1 \, v_1^4 + \frac 18 \lambda_2 \, v_2^4 + \frac 14 \lambda_L \, v_1^2 v_2^2 \,,
\end{equation}
with $\lambda_L = \lambda_3 + \lambda_4 + \lambda_5$.
We briefly summarize here the tree-level conditions leading to the
correct electroweak symmetry breaking while respecting the $Z_2$
symmetry, for details see e.g.~\cite{Ferreira:2019bij} and references
therein.  There are in principle four possible minima\footnote{There is also the possibility of a charge breaking local
  minimum which, however, has a value larger than the correct 
  global electroweak symmetry breaking one \cite{Ferreira:2004yd}.}:
\begin{enumerate}
    \item $v_1^2 = v_2^2 = 0$ ;
    \item $v_1^2 = -\frac{2\mu_1^2}{\lambda_1}, \; v_2^2 = 0$ ;
    \item $v_1^2 = 0, \; v_2^2 = -\frac{2\mu_2^2}{\lambda_2}$ ;
    \item $ v_1^2 = -2 \, \frac{\mu_1^2 \lambda_2 - \mu_2^2 \lambda_L}{\lambda_1\lambda_2 - \lambda_L^2}, \; v_2^2 = -2 \, \frac{\mu_2^2 \lambda_1 - \mu_1^2 \lambda_L}{\lambda_1\lambda_2 - \lambda_L^2}$.
\end{enumerate}
The corresponding minimum should fulfil the conditions: $v_{1,2}^2
\geq 0$, the Hessian
should be positive, i.e. $\det \left[ \frac{\partial^2 V_0}{\partial
    v_1 \partial v_2} \right] > 0$. The potential at minimum 2
should be the lowest of the four: Taking $\lambda_1,\lambda_2 >0$, as required to avoid
unbounded from below directions, one finds beside $\mu^2_1 < 0$ the conditions,
\begin{subequations}
    \label{eq:global_min2}
    \begin{equation}
        \label{eq:global_min2_1}
        \lambda_1 < \lambda_L \frac{\mu_1^2}{\mu_2^2} \quad \text{ if } \quad \mu^2_2 < 0 \text{ and } \lambda_L > 0 \, ,
    \end{equation}    
    \begin{equation}
        \lambda_1 > \lambda_L \frac{\mu_1^2}{\mu_2^2} \quad \text{ if } \quad \mu^2_2 > 0 \text{ and } \lambda_L <  0 \, .
    \end{equation}
\end{subequations}
Moreover, minimum 2 is always the global one if $\lambda_L > 0$ and $\mu^2_2 > 0$.
The requirement, that the potential is bounded from below also implies \cite{Klimenko:1984qx},\footnote{The conditions here listed are only necessary conditions, but sufficient for our purposes. For a more complete set see e.g.~\cite{Kannike:2012pe,Kannike:2016fmd}.}
\begin{subequations}
    \label{eq:bounded_below}
    \begin{equation}
        \label{eq:bounded_below1}
        0 < \sqrt{\lambda_1 \lambda_2} + \lambda_3 + \lambda_4 -|\lambda_5|
    \end{equation}    
    \begin{equation}
        0 < \lambda_3 +  \sqrt{\lambda_1 \lambda_2}  
        \label{eq:bounded_below2}
    \end{equation}
\end{subequations}

It has been shown in \cite{Merle:2015gea} that due to the evolution of the
renormalization group equations (RGEs) the parameters get changed
such that the desired $Z_2$ gets spontaneously broken at high energy
scales. The most relevant RGE in this context is the one for
$\mu^2_2$:
\begin{align}
\label{eq:mu2rge}
Q \frac{d \mu^2_2}{d Q} = \frac{1}{16 \pi^2}\left[ \left(6 \lambda_2 - \frac{3}{2}(g^2_Y+3 g^2_L) + 2 \text{tr}(Y^{\dagger}_N Y_N)\right) \mu^2_2 
+ 2 (2	\lambda_3 + \lambda_4) \mu^2_1 - 4 \sum_{i=1}^3 M^2_{N_i} (Y_N Y^\dagger_N)_{ii}
\right]
\end{align}
where $Q$ denotes the renormalization scale.  In particular, the last
contribution proportional to $M^2_{N_i}$ can drive $\mu^2_2$ to
negative values if the corresponding Yukawa couplings are sizeable and
the larger the ratios $M^2_{N_i}/\mu^2_2$ are.
 
We will investigate to which extent this implies that the $Z_2$
symmetry could be broken at high energies during the thermal evolution
of the Universe. For this, one has to take into account the thermal
contributions to the potential and check if there are indeed parameter
points where the $Z_2$ breaking minimum could potentially arise at a
high temperature $T$.  In such a minimum, also $v_2$ would be non-zero
and we give here the corresponding tree-level mass matrices from which
we calculated the input to the one-loop contributions discussed in the
next section. The mass matrices for the CP-odd, -even and charge
scalars in the basis ($H,\eta$) are given by:
\begin{equation} %\label{eq:}
    \mathcal{M}_R^2 = \left(
        \begin{array}{cc}
            \mu_1^2 + \frac 32 \lambda_1 v_1^2 + \frac 12 \lambda_L v_2^2
                &
            \lambda_L v_1 v_2
            \\
            \lambda_L v_1 v_2
                &
            \mu_2^2 + \frac 32 \lambda_2 v_2^2 + \frac 12 \lambda_L v_1^2
        \end{array}
    \right),
\end{equation}
\begin{equation} %\label{eq:}
    \mathcal{M}_I^2 = \left(
        \begin{array}{cc}
            \mu_1^2 + \frac 12 \lambda_1 v_1^2 + \frac 12 \lambda_S v_2^2
                &
            \lambda_5 v_1 v_2
            \\
            \lambda_5 v_1 v_2
                &
            \mu_2^2 + \frac 12 \lambda_2 v_2^2 + \frac 12 \lambda_S v_1^2
        \end{array}
    \right),
\end{equation}
\begin{equation} %\label{eq:}
    \mathcal{M}_{\pm}^2 = \left(
        \begin{array}{cc}
            \mu_1^2 + \frac 12 \lambda_1 v_1^2 + \frac 12 \lambda_3 v_2^2
                &
            \frac 12 (\lambda_4 + \lambda_5) v_1 v_2
            \\
            \frac 12 (\lambda_4 + \lambda_5) v_1 v_2
                &
            \mu_2^2 + \frac 12 \lambda_2 v_2^2 + \frac 12 \lambda_3 v_1^2
        \end{array}
    \right),
\end{equation}
where the lightest eigenvalues of $\mathcal{M}_I^2$ and
$\mathcal{M}_{\pm}^2$ need to be zero when evaluated at any of the
minima, since they correspond to the Goldstones of electroweak
symmetry breaking. Here, 
we have defined $\lambda_L = \lambda_3 + \lambda_4 +
\lambda_5$ and $\lambda_S = \lambda_3 + \lambda_4 - \lambda_5$.
It is worth
  noticing that in the unbroken $Z_2$ phase $\lambda_4-|\lambda_5| < 0$ when $N$ is heavier than all  scalars  to avoid an electrically charged
  stable dark matter candidate. 
The masses of the electroweak gauge bosons are given by
\begin{equation*} %\label{eq:}
    m_W^2 = \frac 14 g_L^2 ( v_1^2 + v_2^2 ), \qquad m_Z^2 = \frac 14 ( g_L^2 + g_Y^2 ) ( v_1^2 + v_2^2 ), \qquad m_\gamma^2 = 0.
\end{equation*}
The standard model charged fermion masses are simply calculated from
the SM Higgs Yukawa couplings as, $m_f^2 = \frac 12 Y_f^2 v_1^2$ whereas for
$v_2\ne 0$ a tree-level type-I seesaw contribution to the
neutrino masses appears, given by: 
\begin{align*}
M_\nu = \begin{pmatrix}
0 & m_D \\ m_D & M_N 
\end{pmatrix} \quad \text{ with } m_D = \frac{1}{\sqrt{2}} Y_N v_2 \,.
\end{align*}

\section{Effective potential, temperature effects and decoupling of heavy degrees of freedom \label{sect:eff}}

We will combine the one-loop RGEs with the effective potential and
include also the thermal contributions. The region of parameter space,
where potentially a $Z_2$ breaking could occur, is where the
right-handed neutrinos are heavier than the additional scalars and
have sizeable neutrino Yukawa couplings \cite{Merle:2015gea}. The mass
hierarchy between $N_i$ and $\eta$ can be quite large.  This implies
that one has to decouple the heavy states properly as outlined for
example in \cite{Manohar:2020nzp}. We will briefly discuss this after
presenting the thermal contributions. In the following we will neglect
the contributions from SM fermions except for the top quark, as they do
not play any role here.

%\subsection{Zero temperature effective potential}

The effective potential at one-loop and zero temperature is given by
$V_{eff}^{(0)} = V_0 + V_1$, where the temperature-independent
one-loop correction to the potential is the Coleman-Weinberg potential
is given by,
\begin{equation} \label{eq:1loop_pot}
    V_1 = \frac{1}{64\pi^2} \sum\limits_i n_i \, m_i^4 \left( \ln \frac{m_i^2}{Q^2} - c_i \right).
\end{equation}
Here, $m^2_i$ are the field depend mass squares of the various particles and  the $n_i$ count the degrees of freedom and the sign is positive for bosons and negative for fermions:  $n_Z = 3$, $n_W = 6$, $n_\gamma = 2$, $n_{S^0} = 1$, $n_{S^{\pm}} = 2$, $n_t = -12$
 and $n_{N_i} = -2$. $S^0$ and $S^{\pm}$ refers to any neutral or charged scalar.
We take the Landau gauge ($\xi = 0$) and work in the  $\overline{MS}$ scheme \cite{Martin:2001vx,Laine:2016hma}. 
Thus, one has $c_i=3/2$ in case of scalars and fermions and $c_i=5/6$ for the vector bosons.

We note for completeness that \eq{eq:1loop_pot} depends on the choice of the renormalization scale $Q$. To minimize its effect one should use the improved effective potential by using the renormalization group, i.e.
\begin{equation} %\label{eq:}
    \left[ Q \frac{\partial}{\partial Q} + \beta_i \frac{\partial}{\partial \lambda_i} - \gamma_{m_i} m_i^2 \frac{\partial}{\partial m_i^2} - \gamma_{v_i} v_i \frac{\partial}{\partial v_i} \right] V_{eff}^{(0)}(Q,\lambda_i,m^2_i,v_i) = 0.
\end{equation}
The solution is given by $V_{eff}^{(0)}(Q,\lambda_i,m^2_i,v_i) = V_{eff}^{(0)}(\,Q,\,\lambda_i(Q),\,m^2_i(Q),\,v_i(Q) \,)$, with $\lambda_i(Q)$, $m^2_i(Q)$ and $v_i(Q)$ the running parameters \cite{Bando:1992np,Quiros:1999jp}.

%\subsection{Finite temperature effective potential}

The effective potential including finite temperature effects is defined as $V_{eff}^{(T)} = V_0 + V_1 + V_T$ at one-loop level. Here, the leading order temperature-dependent correction is given by,
\begin{equation} \label{eq:T_pot}
    V_T = \frac{T^4}{2\pi^2} \left[ \sum\limits_{i \in bosons} n_i \, J_B\!\left( m_i^2/T^2 \right) + \sum\limits_{i \in fermions} n_i \, J_F\!\left( m_i^2/T^2 \right) \right]\,,
\end{equation}
where the functions $J_{F,B}(y^2)$ come from the evaluation of the thermal loop with the corresponding statistic for bosons or fermions and they are defined as,
\begin{equation} %\label{eq:}
    J_{F,B}(y^2) = \int_0^\infty \!\! dx \, x^2 \ln \left[ 1 \pm e^{-\sqrt{x^2+y^2}} \right] \, ,
\end{equation}
see e.g.~\cite{Laine:2016hma} and references therein.

%\subsection{Resummation and thermal masses}

One should take the thermal mass $m(v_i,T)$ to avoid infrared divergences for $T>>m$. The thermal mass is only important for the zero Matsubara mode \cite{Laine:2016hma}, so only boson masses have to be modified according to $m_i^2(v_i) \, \rightarrow \,\Pi_i^2(v_i,T) \equiv m_i^2 + k_i T^2$.
\begin{eqnarray} %\label{eq:}
    \Pi_{W_L}^2 &=& m_{W_L}^2 + 2 g_L^2 T^2,
    \\
    \Pi_{\mu_1}^2 &=& \mu_1^2 + \left[ \frac 18 g_L^2 + \frac{1}{16}(g_L^2+g_Y^2) + \frac 14 \lambda_1 + \frac{1}{6} \lambda_3 + \frac{1}{12} \lambda_4 + \frac{1}{4} Y_t^2 \right] T^2,
    \\
    \Pi_{\mu_2}^2 &=& \mu_2^2 + \left[ \frac 18 g_L^2 + \frac{1}{16}(g_L^2+g_Y^2) + \frac 14 \lambda_2 + \frac{1}{6} \lambda_3 + \frac{1}{12} \lambda_4 + \frac{1}{24} Tr( Y_N^\dagger Y_N ) \right] T^2,
    \\
    \Pi_{Z_L}^2 &=& \frac 12 \left[ m_Z^2 + \Delta(m_Z^2, T^2) \right] + (g_L^2 + g_Y^2) T^2,
    \\
    \Pi_{\gamma_L}^2 &=& \frac 12 \left[ m_Z^2 - \Delta(m_Z^2, T^2) \right] + (g_L^2 + g_Y^2) T^2,
\end{eqnarray}
with $\Delta^2 = m_Z^4 + (g_L^2-g_Y^2)^2 \, ( 4 T^2 + v_1^2 + v_2^2) \, T^2$. Note, that
in case of the gauge bosons only the longitudinal components get a thermal contribution
from the zero Matsubara mode.

The renormalization scale $Q$ should be chosen such, that large
logarithms are avoided in the calculation of the loop corrected
effective potential in \eq{eq:1loop_pot}.  We use here the
$\overline{MS}$ scheme where one has to decouple heavy states by hand.
In view of the fact that a potential $Z_2$ breaking requires
$M^2_{N_i}/\mu^2_2 \gsim 1$, we have chosen the following procedure:
assuming that the additional scalars have masses in the range of a few
hundred GeV we decouple the right-handed neutrinos if their mass is
above max(1 TeV, $T$). The idea here is that right-handed neutrinos
will only contribute if the temperature $T$ is sufficiently high. The
renormalization scale is then chosen to be:
\begin{align}
\label{eq:Qrelation}
Q = \sqrt[2 n]{\Pi_{k=1}^n m^2_k(v_i,T)} \, ,
\end{align}
where $3 \le n \le 6$ depending on how many neutrinos decoupled, for
instance $n=3$ if all of them are decoupled and 6 if none is
decoupled. In the RGE evolution, as well as in the calculation of the
effective potential including the thermal effects, the right-handed
neutrinos are included if their masses $m_{N_i} \le Q$, otherwise they
are decoupled.  We have checked that the results presented in the next
section remain the same if we replace this geometric mean for $Q$ by
the maximum of the masses which are taken into account.

\section{Numerical results \label{sect:num}}

As already explained in the previous sections, the improvement of the
RGEs to the effective potential implies that the potential only
depends on four scales that can be related: the renormalization
scale $Q$, the temperature $T$ and the background fields values $v_1$
and $v_2$. To keep the logarithmic corrections under control, we
considered the renormalization scale as a function of the thermal mass
\eq{eq:Qrelation}. Then, for each set of input values, we numerically
minimized the effective potential in terms of $v_1$ and $v_2$ for
different values of the temperature ranging from $0$ to $10^{16}$ GeV,
building our way back in time in the evolution of the universe. The
mass spectrum and the one-loop RGEs are numerically computed by a
modified version of \texttt{SPheno} \cite{Porod:2003um, Porod:2011nf}
generated by \texttt{SARAH} \cite{Staub:2013tta,Staub:2015kfa}. 
The minimization of the thermal effective potential is done using
\texttt{MINUIT} \cite{James:1975dr}. Certain checks and cuts are applied during the
procedure regarding unboundedness from below and
non-perturbativity. The initial values should generate a valid
electroweak minimum, i.e. fulfil the conditions
\eq{eq:global_min2}. Also, we only considered points that satisfy the
bounded from below conditions \eq{eq:bounded_below} at every step of
the RGE running. One should be especially cautious about this point, as
the well-known bounded from below conditions of the scotogenic or the
IDM model are just necessary conditions, but not
sufficient.\footnote{We indeed found that from the set of randomly
  generated input values, around a $0.05$\% led to an unbounded from below
  potential even passing the conditions \eq{eq:bounded_below}. Of
  course, such points were excluded from our result.} Then, we checked
possible non-perturbative couplings, as depending on the initial
values at low scale, the RGE running can drive the couplings outside
the perturbative regime. Our computation of the effective potential
cannot be trusted in such cases, so we omitted these points.  

\begin{figure}[tb!]
    \centering
    \includegraphics[width = 0.7\textwidth]{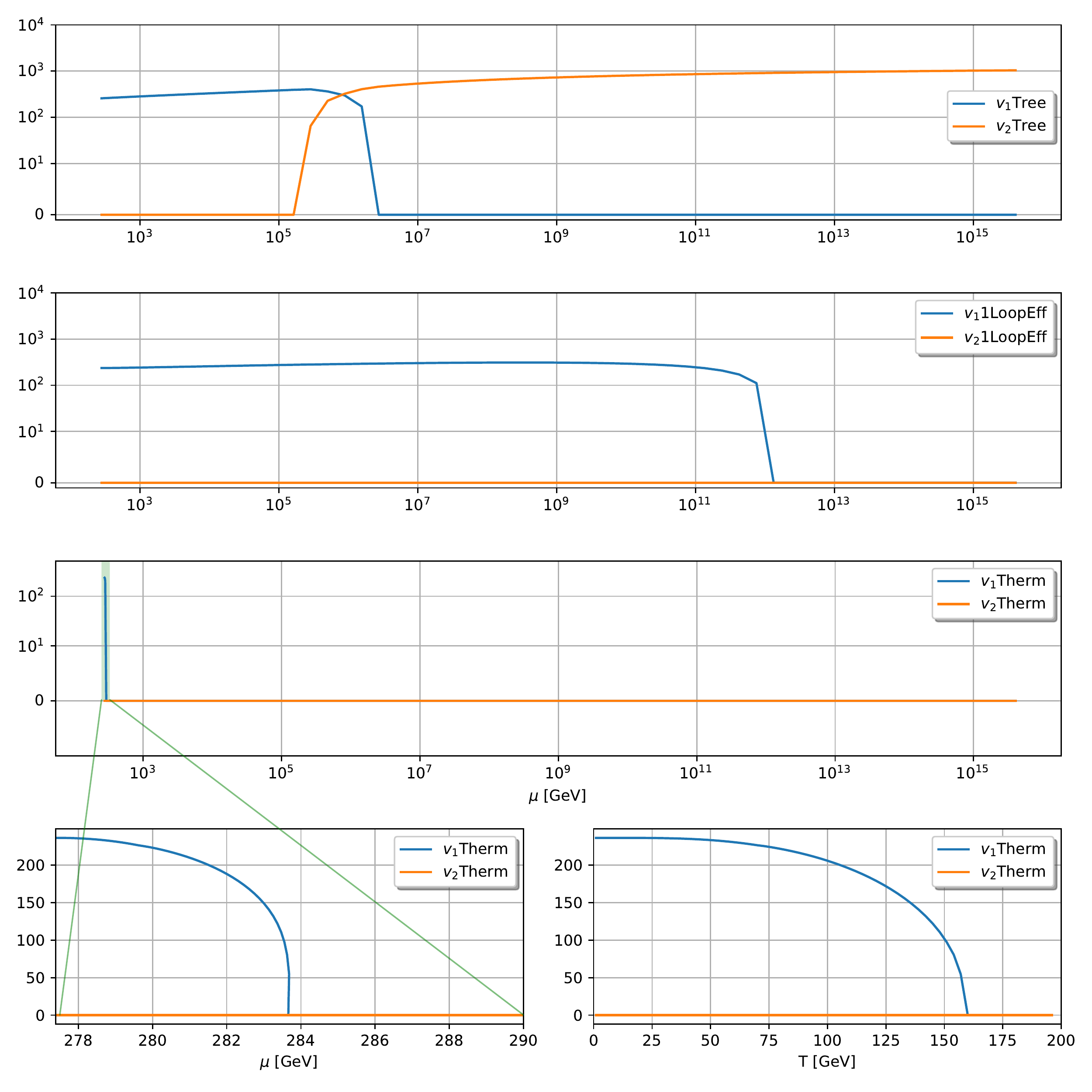}
    \caption{Comparison of the behaviour of the vevs at the minimum
      for a benchmark point for different RGE-improved
      potentials. From top to bottom: tree-level potential (top),
      one-loop corrected potential and the thermal effective
      potential. The last panel zooms into the region where the
      symmetry is restored as a function of the energy scale (left)
      and temperature (right). The values for $v_1$ and $v_2$ are
      given in GeV. Values of the parameters: ($\lambda_2$,
      $\lambda_3$, $\lambda_4$, $Y_n$, $\mu_2^2$, $M_n$) = ($0.2$,
      $0.2$, $-0.15$, $0.2$, $3$ TeV$^2$, $1$ TeV).}
    \label{fig:plotvevs}
\end{figure}

In \fig{fig:plotvevs} we show the behaviour of the $v_1$ and $v_2$
minima for a benchmark point as a function of the energy scale. We
consider three different RGE-improved potentials: tree-level potential
only, one-loop zero-temperature potential and thermal effective
potential. For this point, $\mu_2^2$ runs into negative values around
$10^5$-$10^6$ GeV \eq{eq:mu2rge}. We see that when considering only
the tree-level potential the vev of $\eta$ becomes non-zero around
this scale, when the condition \eq{eq:global_min2_1} fails. This
contrasts with the behaviour of the one-loop potential or the thermal
potential, where no $Z_2$-breaking minimum is found. Once the correct
treatment of the potential with the effective one-loop improved
potential is taken into account, the spontaneous breaking of $Z_2$
arising from the RGEs disappear. For completeness, in the last panel
we show a \textit{zoom} of the region of interest for the last
potential just to compare with the previous case. As stated in
\cite{Katz:2014bha, Dine:1992wr} the symmetry restoration should occur
at temperatures around $150$-$170$ GeV, which is what we found
here. This leads to a merging of the global and local minima at
$v_{1,2}=0$.  

\begin{figure}[tb!]
    \centering
    \includegraphics[width = 0.75\textwidth]{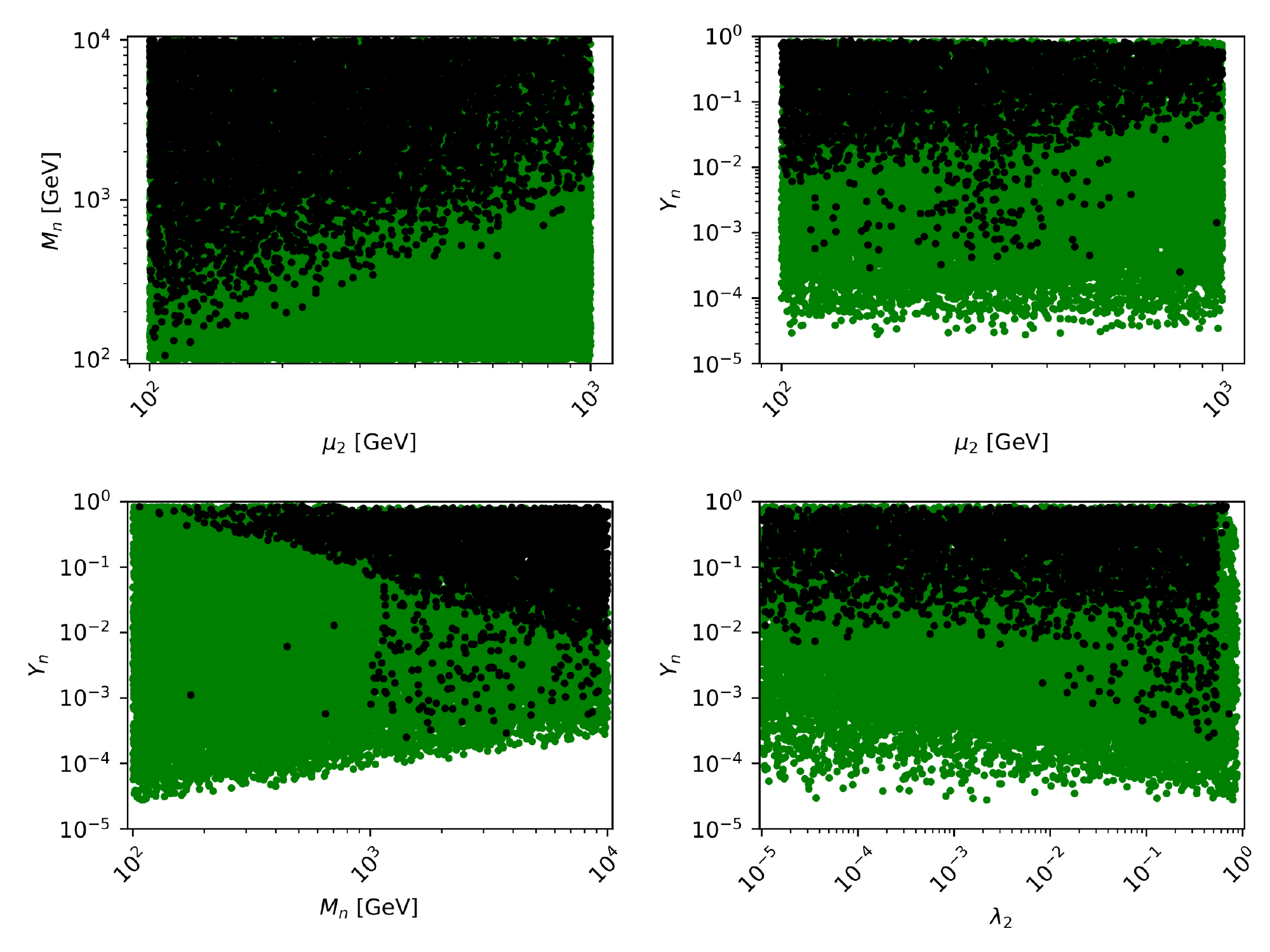}
    \caption{Results of the scan over the parameter space of the
      scotogenic model. No $Z_2$-breaking minimum was found at any step
      of the thermal evolution for the parameter space we
      studied. Green dots represent points which do not break $Z_2$
      symmetry when the effective thermal potential is
      considered. Black dots are $Z_2$-breaking according to
      \cite{Merle:2015gea}, but not in our approach.}
    \label{fig:logscan_Merles}
\end{figure}

We scanned over the input BSM parameters $(Y_N)_{ij}$, $M_{N_k}$,
$\mu_2$ and $\lambda_{2,3,4}$ searching for a $Z_2$ breaking minima at
any step in the thermal evolution. For simplicity, we
considered the mass hierarchy $M_{N_k} = (1, 1.5, 2) \times M_n$, as
well as all the entries of the diagonal Yukawa matrix $Y_N$ to be the same and
equal to $Y_n$.
We randomly generate $5\times 10^4$
points in logarithmic scale in the ranges,
\begin{equation}
    \begin{cases}
    |\lambda_{2,3,4}| :& [10^{-5}, 1]\\
    |Y_n| :& [10^{-5}, 1]\\
    \mu_2 :& [10^2, 10^3]~GeV\\
    M_n :& [10^2, 10^4]~GeV
    \end{cases}
    \label{limit_scan_log} \, .
\end{equation}
Note that $\lambda_5$ is
  not among the input parameters, as the neutrino mass
  \eq{eq:numassSC} can be used to fix its value.
All input parameters are given at the scale $Q=160$~GeV [$\simeq m_t(m_t)$].  As $\lambda_3$ and $\lambda_4$ can be negative, the sign
is also randomized. In the RGE evolution the top-Yukawa coupling $Y_t$ and the strong
coupling $g_3$ are numerically important, in particular for $\lambda_1$. We have set their values at $Q=160$~GeV to $Y_t=0.93761$ and $g_3=1.16787$.
Dimensionless parameters are conservatively taken
up to $1$, as even for this value the running will drive the parameter
to the non-perturbative regime,  $\sqrt{4 \pi}$ in our case,
 at a very low scale. $\mu_2$ should not be very large to
maximize the possibility of getting a $Z_2$ breaking minimum, but
cannot be very low, as there are already constraints to the mass of
$\eta$ around $100$ GeV \cite{ParticleDataGroup:2020ssz}. Regarding
$M_n$, naively one would expect that its effect gets larger the larger it is. 
However, a large mass means that it gets effective at higher temperatures
implying that it contributes only at a later stage to the RGE evolution. 
 We are thus interested in making this value large
to drive $\mu_2$ faster towards negative values when doing the running, see \eq{eq:mu2rge}, but not too large so right-handed neutrinos couple at
sufficiently low energies before the thermal contribution becomes
dominant. For each set of input values we computed the RGE-improved
effective thermal potential and minimized it with respect to $v_1$ and
$v_2$ for different temperatures.  Finally, from the list of minima for each temperature, we
search for $Z_2$ symmetry breaking minima, i.e.\ $|v_2| > 0$ (actually,
larger than the numerical uncertainty of \texttt{MINUIT}). We have considered here
temperatures $T$ up to $10^{16}$ GeV in the cases where all couplings
remain perturbative. We have stopped at lower temperatures if the modulus of one of the quartic 
couplings exceeds $\sqrt{4 \pi}$ as latest at this stage perturbation theory breaks down.
 As can be seen in \fig{fig:logscan_Merles}, we found no $Z_2$ breaking points in
the parameter space considered.

Figure~\ref{fig:logscan_Merles} also shows, for comparison, the points
which break $Z_2$ symmetry according to \cite{Merle:2015gea}. The
black dots are points within our scan that would be considered to be
$Z_2$ symmetry breaking in \cite{Merle:2015gea}, but not in our
approach, using the whole thermal effective potential (see
\fig{fig:plotvevs} and explanations therein). Note that in
\cite{Merle:2015gea} the authors evaluated and minimized only the
tree-level potential at different steps of the RGE running, finding
that the RGEs can drive $\mu_2^2$ towards large negative values
generating a $Z_2$ breaking minimum. This behaviour is clear in
\fig{fig:logscan_Merles}, the negative contribution to the running of
$\mu_2^2$ is proportional to $M_n^2$ and $Y_n^2$, so the black points
appear towards larger values of $M_n$ and $Y_n$, and of course
lower initial values of $\mu_2$.

\section{Conclusions}

In this work we have studied the effects of temperature on the
$Z_2$ symmetry breaking in the scotogenic model. It had previously been
shown \cite{Merle:2015gea} that in the running of the parameters of the
scotogenic model to high energy, the mass squared parameter for the
inert doublet turns negative in sizeable parts of parameter space. 
It was concluded then in \cite{Merle:2015gea} that such points 
should be excluded from the parameter space, since dark matter 
would be lost due to the spontaneous breaking of the $Z_2$ symmetry 
by the vacuum expectation value of $\eta$. 

Here, however, we have shown that this conclusion was premature. 
For a more realistic calculation, we have constructed the effective
potential of the scotogenic model, taking special care on the inclusion
of thermal effects. We then take into account the RGE evolution of
parameters, including the correct decoupling of the right-handed
neutrinos in case that they are heavy. Our numerical scans over the
parameter space then show, that no spontaneous $Z_2$ breaking occurs
in the parts of parameter space studied in \cite{Merle:2015gea}.
Our scans are quite general, but since our numerical programs fail
for points very close to non-perturbativity, we can not make any
claims for that part of parameter space. With this possible caveat,
however, our main result is that considering the complete one-loop 
contribution to the effective potential with the thermal corrections, changes
the conclusion about the stability of dark matter in the early universe.

\bigskip

%%%%%%%%%%%%%%%%%%%%%%%%%%%%%%%%%%%%%%%%%%%%%%%%%%%%%%%%%%%%%%%%%%%%%%%%%%%%%%%%%%%%%%%%%%%%%%%%%%%%%%
% ACKNOWLEDGEMENTS
\centerline{\bf Acknowledgements}

\medskip

Work supported by the Spanish grants PID2020-113775GB-I00
(AEI/10.13039/501100011033) and PROMETEO/2018/165 (Generalitat
Valenciana).  R.C. is supported by the Alexander von Humboldt
Foundation Fellowship. A.A.~and W.P.~are supported by DFG, project
nr. PO-1337/8-1.

\bigskip
 
%%%%%%%%%%%%%%%%%%%%%%%%%%%%%%%%%%%%%%%%%%%%%%%%%%%%%%%%%%%%%%%%%%%%%%%%%%%%%%%%%%%%%%%%%%%%%%%%%%%%%%
% APPENDICES
%\appendix
%\input{appendix}

%%%%%%%%%%%%%%%%%%%%%%%%%%%%%%%%%%%%%%%%%%%%%%%%%%%%%%%%%%%%%%%%%%%%%%%%%%%%%%%%%%%%%%%%%%%%%%%%%%%%%%
% BIBLIOGRAPHY
\bibliographystyle{t1}
\bibliography{bibliography}

\end{document}